\documentclass[]{spie}  

\usepackage{amsfonts}
\usepackage{amsmath}

\usepackage{bbm}
 
\usepackage{amsmath,amsfonts,amssymb}
\usepackage{graphicx}
\usepackage[colorlinks=true, allcolors=blue]{hyperref}

\title{Structure-aware Unsupervised Tagged-to-Cine MRI Synthesis with Self Disentanglement}

\author[a]{Xiaofeng Liu}
\author[a]{Fangxu Xing}
\author[b]{Jerry L. Prince}
\author[c]{Maureen Stone}
\author[a]{Georges El Fakhri}
\author[a]{Jonghye Woo}

\affil[a]{Gordon Center for Medical Imaging, Massachusetts
General Hospital and Harvard Medical School, Boston, MA 02114 USA}
\affil[b]{Deportment of Electrical and Computer Engineering, Johns Hopkins University, Baltimore, MD 21218 USA}
\affil[c]{Deportment of Neural and Pain Sciences, University of Maryland School of Dentistry, Baltimore, MD 21201 USA}

\begin{document} 
\maketitle

\vspace{+5pt}
\begin{abstract}

Cycle reconstruction regularized adversarial training---e.g., CycleGAN, DiscoGAN, and DualGAN---has been widely used for image style transfer with unpaired training data. Several recent works, however, have shown that local distortions are frequent, and structural consistency cannot be guaranteed. Targeting this issue, prior works usually relied on additional segmentation or consistent feature extraction steps that are task-specific. To counter this, this work aims to learn a general add-on structural feature extractor, by explicitly enforcing the structural alignment between an input and its synthesized image. Specifically, we propose a novel input-output image patches self-training scheme to achieve a disentanglement of underlying anatomical structures and imaging modalities. The translator and structure encoder are updated, following an alternating training protocol. In addition, the information w.r.t. imaging modality can be eliminated with an asymmetric adversarial game. We train, validate, and test our network on 1,768, 416, and 1,560 unpaired subject-independent slices of tagged and cine magnetic resonance imaging from a total of twenty healthy subjects, respectively, demonstrating superior performance over competing methods.

\end{abstract}


\vspace{+5pt}
\section{Introduction}

Cross-modality image-to-image translation plays a vital role in medical image analysis. For example, tagged-to-cine magnetic resonance (MR) imaging (MRI) synthesis can potentially be used to reduce the extra cine MRI acquisition time and cost, without interfering with subsequent tasks, such as motion analyses \cite{liu2021dual,liu2021generative}. To this end, previous works usually relied on the paired cine and tagged MRI for training \cite{liu2021dual}. However, because of imaging artifacts or because patients might not tolerate long enough in the scanner, one of the modalities (e.g., cine-MRI) could be missing. In such circumstances, it is necessary to relax the requirement of the paired training data. 

A typical solution for unpaired image-to-image translation would be to use the cycle constraint of the bi-directional mapping---e.g., CycleGAN \cite{zhu2017unpaired}, DiscoGAN\cite{kim2017learning}, and DualGAN\cite{yi2017dualgan}. Although prior works yielded visually realistic results by means of a generative adversarial network (GAN) loss \cite{liu2021adversarial,liu2021recursively,liu2021unified,liu2021subtype,he2020classification}, the structural consistency was not enforced, thereby easily resulting in local distortions in the synthesized images. We note that, in the tagged-to-cine MRI synthesis task, keeping the anatomical structure is essential for the subsequent tissue segmentation and motion analysis. In related developments, a segmentation task was added for co-training \cite{zhang2018translating}. However, it requires additional segmentation labels to train an anatomical segmentation network \cite{liu2020importance,liu2020severity,liu2020wasserstein,liu2021segmentation}. In addition, Yang et al. \cite{yang2020unsupervised} enforced the consistency of the MIND feature between input CT and output MRI data. However, that work cannot be applied to our tagged-to-cine MRI synthesis, since the contour patterns cannot be the same between tagged- and cine-MRI due to tag patterns.   

In this work, to address the aforementioned challenges, we propose to achieve a structure-aware translator, by learning general structural features, compared with task-specific or hand-crafted features, such as the contour-based MIND feature \cite{yang2020unsupervised}. This is achieved by an explicit disentanglement of the anatomical structures (e.g., tongue shape) and imaging modalities via a novel input-output image patches self-training protocol. Our framework can be simply added on top of the conventional cycle-constrained GANs for structure-preserving synthesis.

\begin{figure}[t]
\begin{center}\vspace{+10pt}
\includegraphics[width=1\linewidth]{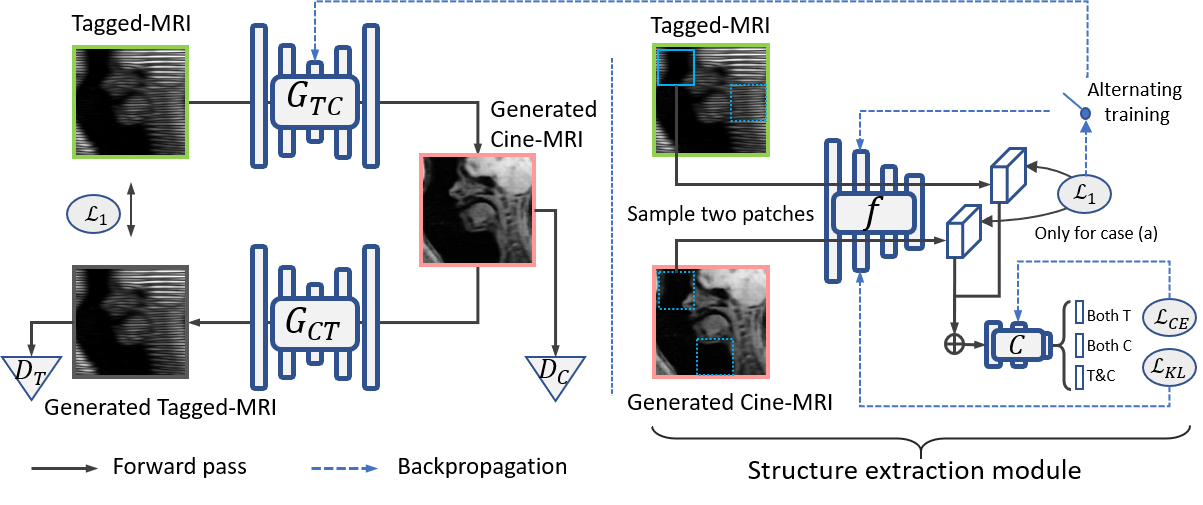}
\end{center} 
\caption{Illustration of our framework for the tagged-to-cine MRI synthesis, which consists of a synthesis module and a structure extraction module.} 
\label{ccc}\end{figure}

\vspace{+5pt} 
\section{Related work}

Tagged MRI has been a crucial imaging modality for measuring tissue deformation in moving organs \cite{fangxu2022spie}. Since it has intrinsically low anatomical resolution, an additional matching set of cine MRI is typically acquired for subsequent tissue segmentation, which added extra scanning time and cost. With the development of GAN-based image generation methods \cite{liu2019hard,liu2020auto3d,liu2021mutualpr,he2020image2audio}, recent methods \cite{liu2021dual,liu2021generative} proposed to synthesize cine MRI from acquired tagged MRI. In these methods, precisely co-registered paired tagged and cine MRI datasets were required for training.

Cycle reconstruction for image style translation is an important technology for the unpaired translation task \cite{zhu2017unpaired,kim2017learning,yi2017dualgan}. However, local structures can hardly be constrained, which thus leads to significant distortions \cite{yang2020unsupervised}. To enforce structural consistency, a recent work~\cite{yang2020unsupervised} incorporated the MIND texture feature to extract an additional supervision signal for CT-to-MRI translation. In addition, deformation invariant CycleGAN\cite{wang2018unsupervised} was proposed to alleviate large nonlinear deformations. Similarly, in that work, a structural dissimilarity loss \cite{xiang2018unpaired} was enforced to preserve local structural consistency. Though low-level texture features, e.g., MIND feature, work efficiently for MR to CT translation, those features cannot be used in our tagged to cine MRI translation task, due to additional tag patterns present in tagged MRI.

 
\vspace{+5pt}
\section{METHODS}

\subsection{MRI Data Acquisition}
All participants were speaking a word, ``a souk,” along with a periodic metronome-like sound during the MRI scan.~MRI scanning was carried out on a Siemens 3.0T TIM Trio system with a 12-channel head coil and a 4-channel neck coil using a segmented gradient echo sequence~\cite{xing2016analysis}. The field of view was 240$\times$240 mm. In-plane resolution was 1.87$\times$1.87 mm and a slice thickness was 6 mm. The image sequence was acquired at the rate of 26 fps. We note that both cine and tagged MRI are in the same coordinate space. The detailed collection protocol can be found in \cite{liu2021dual}. 

\subsection{Our Proposed Network}

In an unpaired setting, we have a group of tagged MR images $\{x_t\}$ and a group of cine MR images $\{x_c\}$. The basic framework of CycleGAN \cite{zhu2017unpaired} is illustrated in Fig. \ref{ccc} left, which has two bi-directional U-Net-based translators $G_{TC}$: Tagged MRI$\rightarrow$Cine MRI and $G_{CT}$: Cine MRI$\rightarrow$Tagged MRI, and two discriminators $D_T$ and $D_C$ for tagged and cine MRI, respectively. In addition, the cycle constraint minimizes the image-level reconstruction loss of $L_1^I$. Note that we only focus on the tagged-to-cine MRI synthesis in this work. Specifically, the optimization objectives can be formulated as:
\begin{align}
    &L_1^I =||G_{CT}(G_{TC}(x_t))-x_t||_1,\\
    &L_{D_C}= \mathbb{E}_{x_c}[\text{log}(D_C(x_c))] + \mathbb{E}_{x_t}[\text{log}(1-D_C(G_{TC}(x_t))],\\
    &L_{D_T}= \mathbb{E}_{x_t}[\text{log}(D_T(x_t))] + \mathbb{E}_{x_c}[\text{log}(1-D_T(G_{CT}(G_{TC}(x_c)))],
\end{align}
which are trained in a round-based manner in each iteration.

A recent study \cite{zhang2018translating} demonstrated that with a bijective geometric transformation $T$ with its inverse transformation $T^{-1}$, the cycle constraint can still be satisfied with the translators of $G_{TC}'=G_{TC}T$ and $G_{CT}'=G_{CT}T^{-1}$. Therefore, the geometrical distortions were not punished at the training stage. In addition, Yang et al. \cite{yang2020unsupervised} extracted the MIND feature $M(x_t)$ and $M(G_{TC}(x_t))$ of $x_t$ and $G_{TC}(x_t)$ with a manually defined extractor $M$, and minimized their reconstruction loss $||M(x_t)-M(G_{TC}(x_t))||_1$. Specifically, since the MIND feature mainly focuses on the boundaries, it it not straightforward to apply the same MIND feature to tagged MRI due to tag patterns. To improve upon the prior work~\cite{yang2020unsupervised}, in the present work, we propose to learn a general structure feature extractor $f$ as an alternative to $M$. Our $f$ can be learned with a novel input-output image patches self-training scheme to achieve a disentanglement of the structures and imaging modalities. 

Inspired by the recent self-supervised learning \cite{chen2019self}, we simply split both the input $x_t$ and the output $G_{TC}(x_t)$ to a set of $3\times3$ patches. In each forward pass, we randomly choose two patches from a batch of 18 patches as our input. Accordingly, there are four possible combinations: 

(a) From the same position of tagged MRI $x_t$ and generated cine MRI $G_{TC}(x_t)$.

(b) Both from tagged MRI $x_t$, with different positions.

(c) Both from generated cine MRI $G_{TC}(x_t)$, with different positions.

(d) From different modalities and different positions.

Based on these four cases, we have two assumptions: (1) the patches in the case of (a) should have consistent anatomical structures, and (2) the patch pairs in these cases have three combinations from two modalities---i.e., both from tagged MRI, both from cine MRI, and from tagged and cine MRI.

To remove the modality information from the embedding of $f$, we adopt a classifier $C$ for checking the modality combination of the concatenated two patch features. $C$ is trained to minimize a 3-class cross-entropy loss $\mathcal{L}_{CE}$ of modality combination classification as the adversarial classifier as in Pix2Pix \cite{isola2017image}. There can be three possible binary combinations, i.e., $\{f(G_{TC}(x_t)),f(x_t)\}$, $\{f(G_{TC}(x_t)),f(G_{TC}(x_t))\}$, and $\{f(x_t),f(x_t)\}$. We denote the true binary combination as $y\in\mathbb{R}^3$, which is a one-hot vector. The combination classification loss can be formulated as:
\begin{align}
    \mathcal{L}_{CE}= \sum_i^3 -y_i\log C_i(\cdot), 
\end{align}
where $C_i(\cdot)$ is the classifier prediction for the i-th class. Our classifier $C$ and feature extractor $f$ play an asymmetrical adversarial game to encourage that $f$ eliminates the modality information \cite{liu2021mutual}. Rather than maximizing the cross-entropy loss, $f$ minimizes the KL-divergence of its softmax prediction and a uniform distribution. Specifically, we minimize the following loss:
\begin{align}
    \mathcal{L}_{KL}=D_{KL}(C(\cdot)||\mathcal{U}).
\end{align}
We note that the modality and position label of the sampled two patches are known, which can be used for supervised training.

\begin{figure}[t]
\begin{center}
\includegraphics[width=1\linewidth]{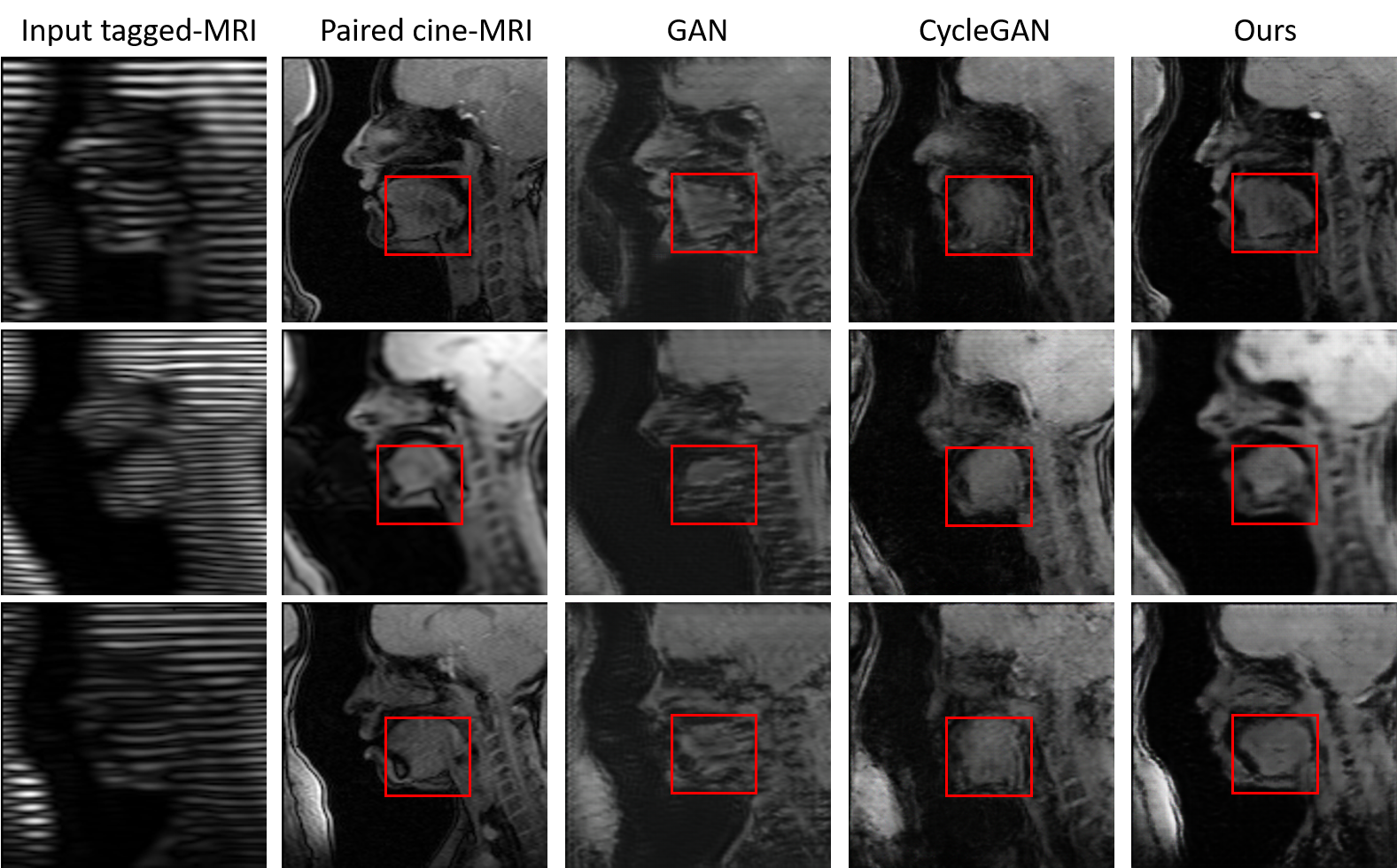} 
\end{center} 
\caption{Comparison of different unpaired tagged-to-cine MR generation methods, including the vanilla GAN (i.e., the half direction of CycleGAN), CycleGAN~\cite{zhu2017unpaired}, and our proposed method. * indicates the first attempt at unpaired tagged-to-cine MR image synthesis.} 
\label{fig:results}\end{figure} 

Other than removing the modality information from the embedding of $f$, it is necessary to preserve the anatomic information. In the case of (a), we would expect the $\mathcal{L}_1^f$ loss between the embeddings of two patches can be minimized. 
\begin{align}
    &\mathcal{L}_1^f =||f(G_{TC}(x_t))-f(x_t)||_1,
\end{align}

Similar to the self-training \cite{liu2021generative}, we adopt an alternating training scheme, which firstly fixes the structure extraction module to update $G_{TC}$ with the $\mathcal{L}_1^f$ loss, and then fix $G_{TC}$ to update the structure extraction module with the $\mathcal{L}_1^f$ loss.

After several iterations, the structure extraction module can be expected to embed the anatomic information and filter the modality information out, and the CycleGAN can be well regularized by the $\mathcal{L}_1$ loss---i.e., the structural consistency constraint. We use $f$ to replace the manually defined $M$ as in \cite{yang2020unsupervised} to achieve a better structural consistency translation. In testing, only the trained CycleGAN part is used for translation.

 
\section{RESULTS}

During training, we used unpaired 1,500 tagged MR images with horizontal tag patterns and 1,500 cine MR images from a total of ten subjects. In addition, two subjects (416 slice pairs) and eight subjects (1,560 slice pairs) are used for hyper-parameter validation and evaluation, respectively. 

For a fair comparison, we resized the tagged and cine MR images to 256$\times$256 and adopted the $G_{TC}$, $G_{CT}$ backbones from CycleGAN \cite{zhu2017unpaired}. For the structure encoder $f$, we used five fully convolutional layers, which resulted in features with the size of $32\times32\times128$. The classifier $C$ has two convolutional layers and two fully-connected layers with a three-dimensional output layer. We used the PyTorch deep learning toolbox for our implementation. We also used an NVIDIA V100 GPU for the training, which took about 6 hours for 200 epochs.

The synthesis results using GAN without the cycle constraint, CycleGAN \cite{zhu2017unpaired}, and our proposed method are shown in Fig.~\ref{fig:results}. We can see that the CycleGAN produced realistic images, while not being able to achieve structural consistency. By contrast, our method was able to keep the position and shape of the tongue consistent between the two modalities. The resulting images were expected to have realistic appearances and to be structurally consistent with its corresponding paired ground-truth $x_c$. For our quantitative evaluation, we used four evaluation metrics, including mean L1 error, structural similarity index measure (SSIM), peak signal-to-noise ratio (PSNR), and inception score (IS) \cite{liu2021dual}. Table \ref{tabel:1} lists numerical comparisons for the eight testing subjects. The proposed framework outperformed the other two comparison methods w.r.t. SSIM, PSNR, and IS. Of note, all of the compared methods do not have the L1 minimization objective as in \cite{liu2021dual}.



\vspace{+5pt}
\section{CONCLUSION}

In this work, we proposed a novel input-output image patches self-training scheme to achieve a disentanglement of the anatomical structures and imaging modalities. The structure extraction module and the structure tagged-to-cine MRI translator $G_{TC}$ were trained with an alternating training protocol. Our translator was able to achieve the structure-aware translation as demonstrated by the tagged-to-cine MRI synthesis task. Both qualitative and quantitative evaluation results showed that our framework outperformed CycleGAN for unpaired training. Our approach can be a simple add-on module to CycleGAN, or the other cycle constrained translators---e.g., DiscoGAN and DualGAN. Additionally, our method can be applied to other modality synthesis tasks, including MRI-to-CT synthesis, which is subject to our future work.

\begin{table}[t]
\centering
\caption{Numerical comparisons of four methods in testing across 1,560 slice pairs. The best results are in \textbf{bold}.} \vspace{+10pt}
\resizebox{0.7\linewidth}{!}{
\begin{tabular}{c|c|c|c|cccc|ccc}
\hline
Methods& L1~$\downarrow$& SSIM~$\uparrow$ &	PSNR~$\uparrow$ 	&  IS~$\uparrow$\\\hline\hline
GAN         & {183.8}$\pm$0.2&	0.8863$\pm$0.0015          &	25.84$\pm$0.07       &	7.19$\pm$0.14\\
CycleGAN	&174.2$\pm$0.3         &	0.9014$\pm$0.0017          &	27.33$\pm$0.06       &	8.86$\pm$0.13\\
Proposed    &\textbf{168.5}$\pm$0.3&	\textbf{0.9237}$\pm$0.0014 & \textbf{29.52}$\pm$0.08 &  \textbf{9.91}$\pm$0.16\\\hline
\end{tabular}
}
\label{tabel:1}
\end{table}

\vspace{+5pt}
\acknowledgments 

This work is supported by NIH R01DC014717, R01DC018511, and R01CA133015.

\vspace{+5pt}
\bibliography{main} 
\bibliographystyle{spiebib} 

\end{document}